%% file: main.tex
\documentclass[10pt,twocolumn,letterpaper]{article}

\usepackage{iccv}              
\usepackage[accsupp]{axessibility}


\definecolor{iccvblue}{rgb}{0.21,0.49,0.74}
\usepackage[pagebackref,breaklinks,colorlinks,allcolors=iccvblue]{hyperref}

\usepackage{booktabs}
\usepackage{multirow}
\usepackage{comment}
\usepackage{lipsum}
\usepackage{xcolor}
\usepackage{colortbl}
\usepackage{makecell}
\usepackage{caption}
\usepackage{overpic}
\usepackage{newtxtext,newtxmath} 

\def\paperID{10865} 
\def\confName{ICCV}
\def\confYear{2025}

\title{Occlusion-robust Stylization for Drawing-based 3D Animation}

\author{Sunjae Yoon \hspace{0.5cm} Gwanhyeong Koo \hspace{0.5cm} Younghwan Lee \hspace{0.5cm} Ji Woo Hong \hspace{0.5cm} Chang D. Yoo\\
School of Electrical Engineering, KAIST\\
{\tt\small \{sunjae.yoon,kookie,youngh2,jiwoohong93,cd\_yoo\}@kaist.ac.kr}
}

\begin{document}
\maketitle
\input{sec/0_abstract}    
\input{sec/1_intro}
\input{sec/2_method}
\input{sec/3_experiment}
{
    \small
    \bibliographystyle{ieeenat_fullname}
    \bibliography{main}
}

\end{document}

%% file: sec/0_abstract.tex
\begin{abstract}
3D animation aims to generate a 3D animated video from an input image and a target 3D motion sequence. Recent advances in image-to-3D models enable the creation of animations directly from user-hand drawings. Distinguished from conventional 3D animation, drawing-based 3D animation is crucial to preserve artist's unique style properties, such as rough contours and distinct stroke patterns. However, recent methods still exhibit quality deterioration in style properties, especially under occlusions caused by overlapping body parts, leading to contour flickering and stroke blurring. This occurs due to a `stylization pose gap' between training and inference in stylization networks designed to preserve drawing styles in drawing-based 3D animation systems. The stylization pose gap denotes that input target poses used to train the stylization network are always in occlusion-free poses, while target poses encountered in an inference include diverse occlusions under dynamic motions. To this end, we propose Occlusion-robust Stylization Framework (OSF) for drawing-based 3D animation. We found that while employing object's edge can be effective input prior for guiding stylization, it becomes notably inaccurate when occlusions occur at inference. Thus, our proposed OSF provides occlusion-robust edge guidance for stylization network using optical flow, ensuring a consistent stylization even under occlusions. Furthermore, OSF operates in a single run instead of the previous two-stage method, achieving 2.4$\times$ faster inference and 2.1$\times$ less memory. 
%
Project is available at: {\href{https://dbstjswo505.github.io/Drawing-based-3D-Animation-page}{\nolinkurl{github.io/Drawing-based-3D-Animation-page}.}}
\end{abstract}

%% file: sec/1_intro.tex
\section{Introduction}
\label{sec:intro}

\begin{figure}[t]
  \centering
   \begin{overpic}[width=\linewidth]{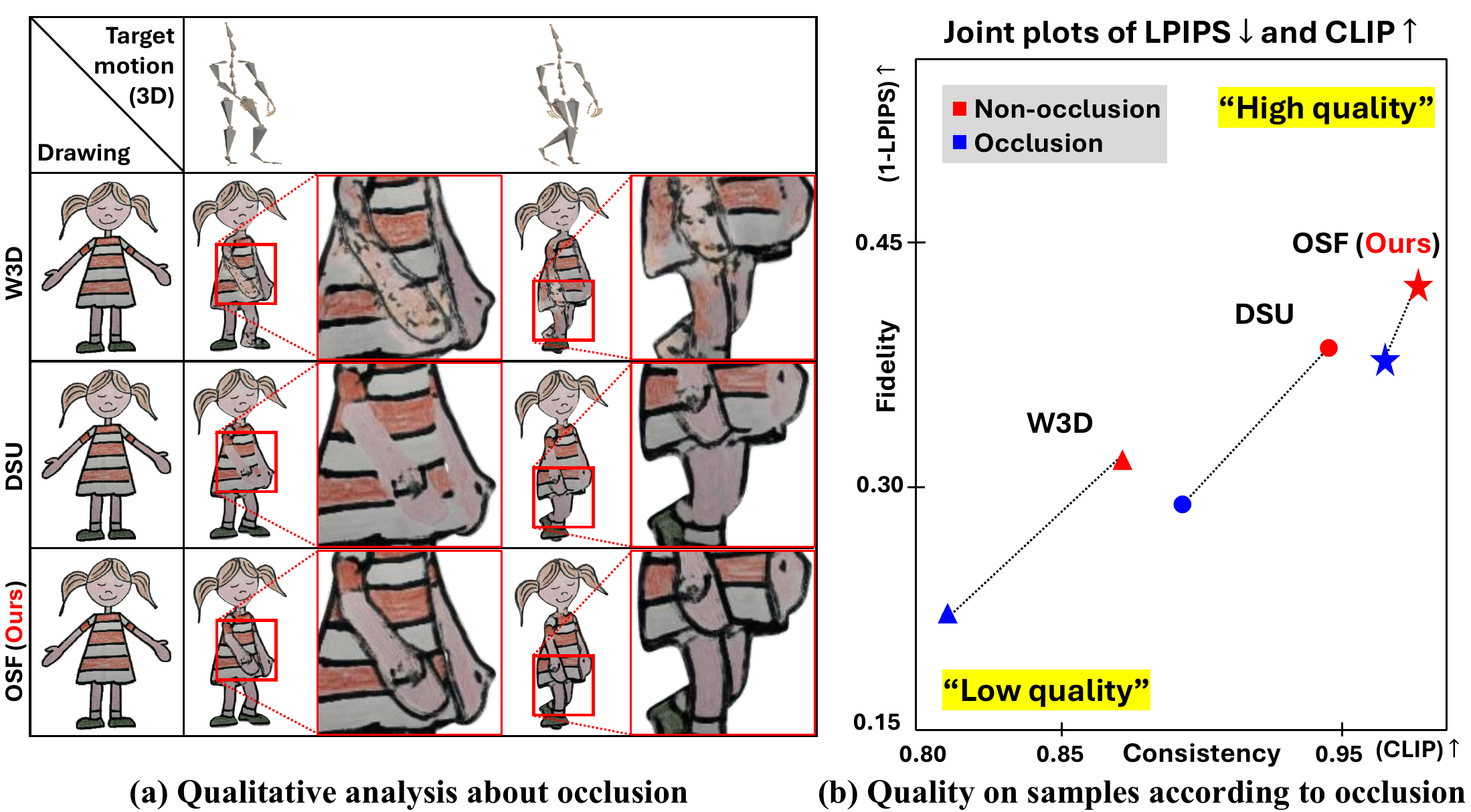}
    \put(0, 26){\rotatebox{90}{\tiny \cite{zhou2024drawingspinup}}}
     \put(0, 39){\rotatebox{90}{\tiny \cite{long2024wonder3d}}}
    \put(75, 26){\tiny \cite{long2024wonder3d}}
    \put(88.7, 33.5){\tiny \cite{zhou2024drawingspinup}}
    \end{overpic}
   \caption{Quality deterioration in occlusion areas: (a) current systems exhibit stylistic inaccuracies in strokes and contours within overlapping body parts, and (b) animations with occlusion reduced consistency (CLIP \cite{radford2021learning}) and fidelity (LPIPS \cite{zhang2018unreasonable}) compared to ones without occlusion across diverse drawings in Amateur Drawings \cite{10.1145/3592788}. The CLIP measures consecutive frame similarity, and LPIPS measures perceptual differences between input and output.}
   \label{fig:1}
   \vskip -0.2in
\end{figure}

%
Denoising diffusion models \cite{song2020denoising,ho2020denoising,song2020score,dhariwal2021diffusion} have reshaped the landscape of generative AI, leading to remarkable advancements \cite{koo2024flexiedit,hong2025ita,kong2020diffwave,ton2025taro,pham2024mdsgen,yoon2024dni} in image, speech, and video generation.
%
Image-to-3D diffusion models \cite{liu2023syncdreamer,liu2023zero,long2024wonder3d} have showcased the potential for creating 3D animations \cite{zhou2024drawingspinup,peng2024charactergen,luo2023rabit} from a single image.
In this work, we focus here on 3D animation techniques for user-drawn images, referred to as drawing-based 3D animation.
%
As shown in Figure \ref{fig:1} (a), drawing-based 3D animation systems take a reference hand-drawn image and a target 3D motion as inputs, and generate 3D animated drawings aligned with the target motion.
In these systems, it is crucial to preserve artists' unique styles, such as contours and stroke patterns in output frames. 
However, synthesizing 3D structures inevitably causes undesirable distortions\footnote{The contour lines depicting the subject’s shape become excessively thick, or the strokes conveying the artist’s unique style become blurred.} on them.
Thus, drawing-based 3D animation systems employ a stylization process using a neural network (referred to as a stylization network) fine-tuned on the input drawing to restore the distortions.
%

\begin{figure*}[t]
  \centering
   \includegraphics[width=\linewidth]{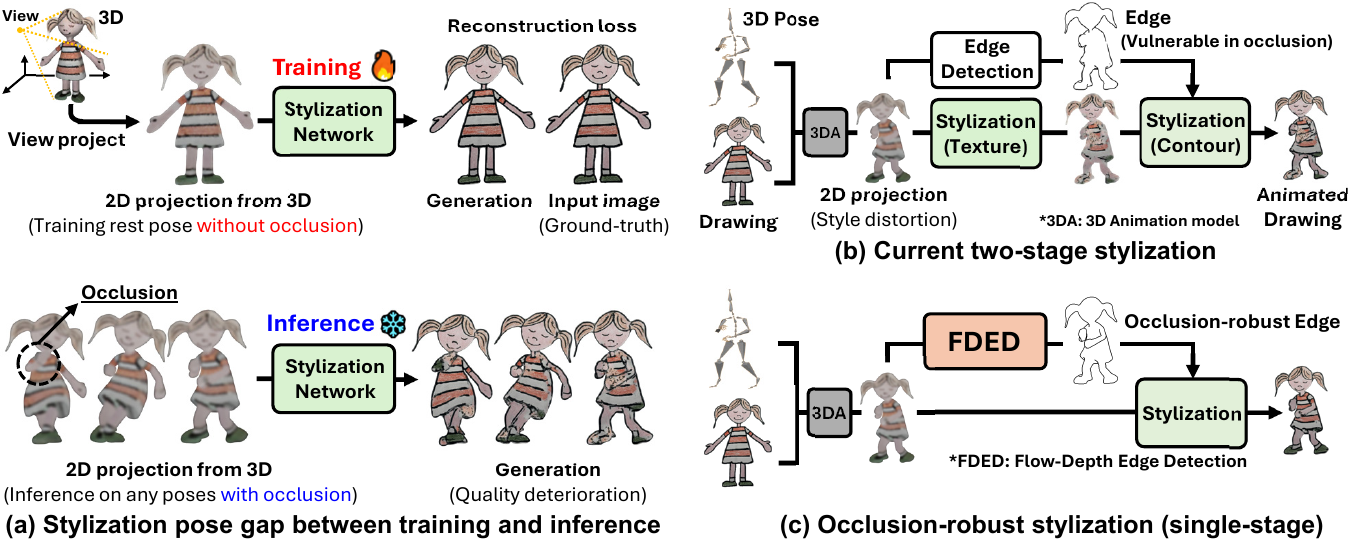}

   \caption{(a) Illustration of stylization pose gap between training and inference. The stylization network is fine-tuned on a single pose (simple occlusion-free pose), due to the unavailability of drawings about the same object's other poses. During inference, the network encounters diverse occluded poses as unseen scenario, reducing stylization robustness. (b) Current two-stage stylization with edge guidance (vulnerable in occlusion) \cite{zhou2024drawingspinup,long2024wonder3d,tang2023dreamgaussian}. (c) Our proposed single-stage occlusion-robust stylization with flow-depth edge detection (FDED).}
   \label{fig:introduction2}
   \vskip -0.1in
\end{figure*}
%
%
Recent advancements \cite{zhou2024drawingspinup,long2024wonder3d,tang2023dreamgaussian} of drawing-based 3D animation systems have demonstrated notable precision in preserving a drawing’s identity within 3D animations.
Nonetheless, these systems still suffer from quality deterioration in terms of drawing properties (\ie contour and stroke) when dealing with occlusions in target poses.
To be specific, in target poses where the object's body parts overlap, the outputs exhibit flickering contours and blurred internal stroke patterns.
Figure \ref{fig:1} (a) presents examples of this issue: when the arm overlaps with the body, the arm's texture and the surrounding clothing of the body are blurred (first row), and the contour line between overlapping parts intermittently appears and disappears (second row) as flickers.
Furthermore, Figure \ref{fig:1} (b) presents quantitative evaluations of the generated animations in terms of consistency and fidelity for target motions with and without occlusion. 
Notably, motions with occlusion yield significantly lower quality compared to those without occlusion.

In fact, this quality deterioration stems from a `stylization pose gap' between training and inference in the stylization network.
The stylization pose gap refers to differences of the input target poses between training and inference of the stylization network.
To be specific, shown in Figure \ref{fig:introduction2} (a), the stylization network is fine-tuned (trained) to restore the original input drawing style onto a 2D projection of 3D animation\footnote{For details of 3D animation, 3D diffusion model \cite{long2024wonder3d} builds 3D structure \cite{mildenhall2021nerf} and apply rigging \cite{baran2007automatic} it on a skeleton to follow 3D target motion.}.
Here, the input drawing serves as ground-truth for the training, where its pose is relatively simple and free from occlusion to show the object's entire body (\eg rest pose in Figure \ref{fig:introduction2} (a)).
During inference, the stylization is applied to diverse poses, including those with occlusions. 
However, the network has not been trained to handle these occluded scenarios because ground-truth drawings for those conditions are unavailable\footnote{We assume a case that only a single ground-truth drawing is available, as obtaining multiple drawings of the same object is not always feasible.}, consequently making it vulnerable in maintaining robust stylization under occlusions.
%
%
\begin{figure}[t]
  \centering
   \includegraphics[width=\linewidth]{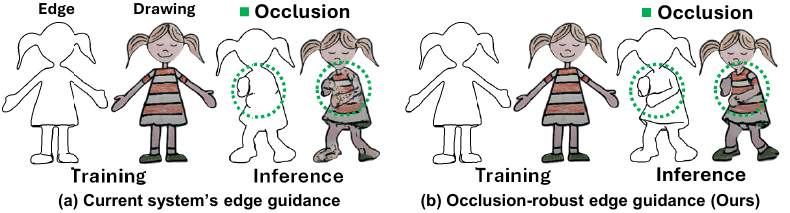}

   \caption{Resulting animated drawings and edges in training and inference: (a) current system's edge and (b) occlusion-robust edge.}
   \label{fig:3}
   \vskip -0.25in
\end{figure}

To this end, we propose  Occlusion-robust Stylization Framework (OSF) to bridge the stylization pose gap between training and inference.
Figure \ref{fig:introduction2} (b) shows the current stylization framework for drawing-based 3D animation system.
Given an input drawing and target 3D pose, 3D animation model (\ie image-to-3D diffusion \cite{long2024wonder3d} and 3D rigging \cite{baran2007automatic}) generates 3D structure following target pose, which is then projected into 2D for animation. 
Since the 2D projection exhibits style distortion, the stylization is performed on it.
Here, edges extracted from the 2D projection are also used as auxiliary input, serving as an effective guideline for the stylization.
As shown in Figure \ref{fig:3}, we observed that the occlusion-free pose (\ie rest pose) used in a training provides clear edges.
However, at inference, occlusions across diverse poses obscure these edges, causing flickering and blurring within the output.
To counter this, our proposed OSF in Figure \ref{fig:introduction2} (c) incorporates flow-depth edge detection, which produces occlusion-robust edges to ensure consistent clarity and robust stylization under occlusion by recovering undetected edges using optical flow.
%
%
Furthermore, current methods rely on a two-stage stylization process (texture to contour), resulting in unnecessary computational overhead and resource usage.
In contrast, our framework operates in a single pass through edge-guided contrastive learning, achieving 2.4$\times$ faster inference and 2.1$\times$ less memory.
\begin{figure*}[t]
  \centering
   \includegraphics[width=\linewidth]{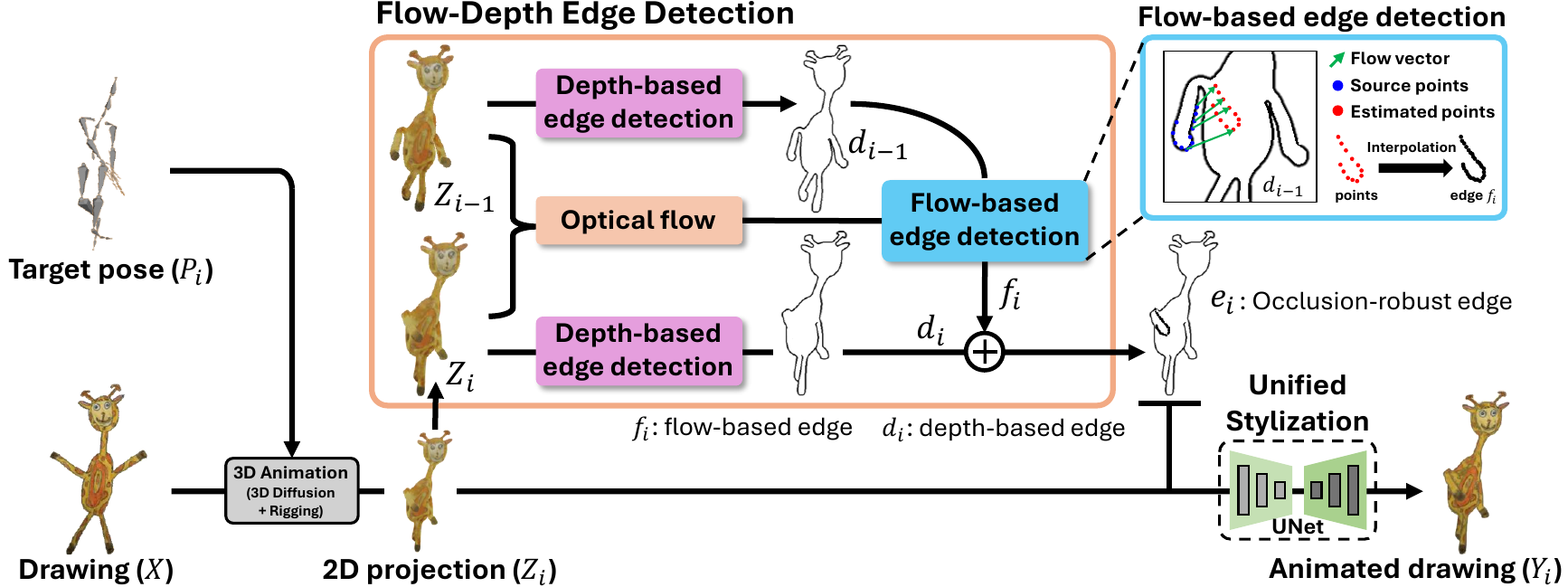}

   \caption{Occlusion-robust Stylization Framework (OSF) for drawing-based 3D animation. A 3D animation model processes a drawing $X$ with the $i$-th target pose $P_i$ to generate 3D structure following the target pose, which is rendered as a 2D projection $Z_i$ for the animation. The proposed OSF stylizes the projection $Z_i$ to final drawing animation $Y_i$, incorporating Flow-Depth Edge Detection (FDED), which produces an occlusion-robust edge $e_i$ to guide robust stylization under unseen occluded poses. The edge $e_i$ combines a depth-based edge $d_i$ for unoccluded region and a flow-based edge $f_i$ for occluded areas by reconstructing missing edges. Our method operates in a single stage, using a simple UNet-based \cite{ronneberger2015u} unified stylization network with our designed edge-guided contrastive learning.
}
   \label{fig:model}
   \vskip -0.15in
\end{figure*}
%
%
\section{Related Works}
\paragraph{Image to 3D Animation.}
Image animation is a temporal sequence of shape deformations designed to ensure seamless transitions.
Early works \cite{hornung2007character,aberman2019learning,albahar2021pose} investigated image-to-image transformations across poses and viewpoints, aided by geometric deformation \cite{sorkine2004laplacian,koo2025flowdrag} and synthesis \cite{cohen2004variational,muller2006procedural}.
Recently, 2D methods \cite{xu2024magicanimate,wang2024disco,siarohin2019first,yoon2024tpc,yoon2024frag} have animated various objects but struggled to capture depth and perspective, prompting a shift to 3D animation \cite{long2024wonder3d,tang2023dreamgaussian,karthikeyan2024avatarone}.
Classic methods constructed 3D meshes from silhouettes and skeletons \cite{buchanan2013automatic,nealen2007fibermesh,schmidt2007shapeshop}, while newer diffusion-based techniques \cite{long2024wonder3d,liu2023syncdreamer} generate multi-view images for optimized 3D reconstructions \cite{izadi2011kinectfusion,mildenhall2021nerf,kerbl20233d}. Rigging\footnote{Creating a skeleton for 3D mesh to control its movements.} process \cite{baran2007automatic,kavan2007skinning} then enabled animation of these meshes according to target motions.
The 3D animations have been applied to diverse images, including human and animal images. 
However, hand-drawn images remain challenging: 3D augmentation distorts contours, and 2D projections (\ie rendering) often cause blurry textures, especially under occlusion.
To address this, we propose a robust stylization framework to maintain consistent quality under diverse occluded motions.

\vspace{-10pt}
\paragraph{Edge Detection in 2D and 3D.}
Edge detection has been a fundamental building block for various computational 2D and 3D visions.
In 2D, conventional methods (\eg Sobel, Canny, and Laplacian) detect boundaries by analyzing gradient changes \cite{canny1986computational}, while recent deep-learning approaches (\eg Holistically-Nested Edge Detection \cite{xie2015holistically}) leverage hierarchical feature extraction for more accurate edge maps. 
Recent works have extended upto 3D data and predict geometric boundary using 3D contours \cite{10.1145/882262.882354}, ridges \cite{10.1145/1276377.1276401} and neural network \cite{liu2020neural}. 
However, these edge detections rely on single frames, making them vulnerable to occlusions. 
By incorporating optical flow in consecutive frames, we provide more robust edge guidance for stylization in occlusion.

%% file: sec/2_method.tex
\begin{figure*}[t!]
  \centering
   \includegraphics[width=\linewidth]{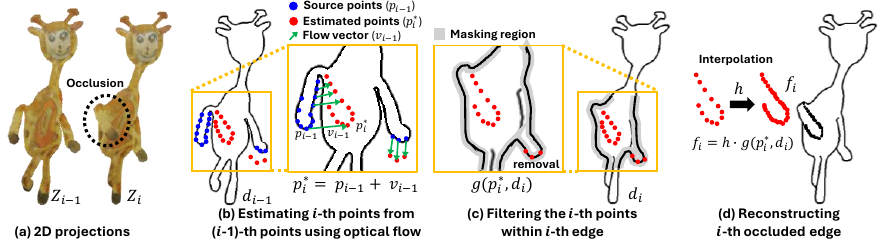}

   \caption{Illustration of flow-based edge detection. (a) shows sequential 2D projections $Z_{i-1}$ and $Z_{i}$ to compute optical flow $v_{i-1}$. (b) shows that the $v_{i-1}$ is added to source points $p_{i-1}$ to estimate their positions $p_{i}^{*}$ in the next $i$-th frame. (c) shows that the points $p_{i}^{*}$ are filtered into internal region of $i$-th edge. (d) shows that the remained points are interpolated to build edges. The $g$ is the filtering operation to filter out estimated points $p^{*}_{i}$ within the region of edge $d_{i}$ and $h$ is interpolation that builds flow-based edge $f_{i}$ from the filtered points.}
   \label{fig:model3}
   \vskip -0.2in
\end{figure*}
\vspace{-8pt}
\section{Method}
Occlusion-robust Stylization Framework (OSF) in Figure \ref{fig:model} is proposed for drawing-based 3D animation.
The 3D animation model (\ie image-to-3D diffusion \cite{long2024wonder3d} with 3D rigging \cite{baran2007automatic}) takes a drawing $X$ and the $i$-th target 3D pose $P_{i}$ to generate 3D animation, rendered as a 2D projection $Z_{i}$ for visualizing it from user's view point. 
3D synthesis and animation often cause style distortions, and stylization addresses these issues in a lower-dimensional space (2D projection) as an intuitive solution.
Formally, stylization takes the $Z_{i}$ as input and predicts the animated drawing $Y_{i}$, where OSF provides occlusion-robust edge guidance $e_{i}$ to enhance stylization robustness under occlusions as below:
\begin{equation}
Y_{i} = F(Z_{i},e_{i}),
\label{eq:stylenet}
\end{equation}
where $F$ is a simple encoder-decoder network (\eg U-Net \cite{ronneberger2015u}).
The edge $e_i$ can be utilized through various operations (\eg concatenation, attention) within the network.\footnote{We adopt a simple early fusion by concatenating it with $Z_i$, but it can be further enhanced with transformer-based interactions \cite{kim2021structured,luu2025enhancing,yoon2023hear,yoon2022information}.}
We focus here on constructing the edge $e$ (For simplicity, we omit subscript $i$).
To be specific, OSF incorporates our designed Flow-Depth Edge Detection (FDED) to generate the occlusion-robust edge $e$. 
This edge can be separated into two parts: edge in occluded regions ($e^{o}$) and the other edge in unoccluded regions ($e^{u}$) as given below:
\begin{equation}
e = e^{o} \cup e^{u}.
\end{equation}
To estimate the two edges, we propose depth-based edge detection for $e^u$ and flow-based edge detection for $e^o$.
%
\begin{figure}[t!]
  \centering
   \includegraphics[width=\linewidth]{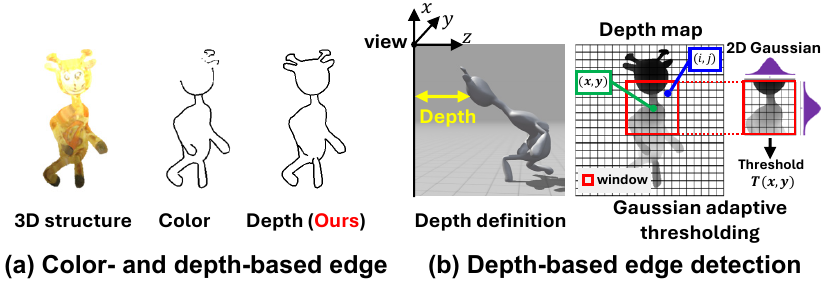}

   \caption{(a) shows the results of color- and depth-based edges with a light source placed on the head of a 3D object. (b) shows depth-based edge detection using Gaussian adaptive thresholding.}
   \label{fig:model2}
   \vskip -0.15in
\end{figure}
\subsection{Depth-based edge detection}
Depth-based edge detection aims to capture the unoccluded edge $e^{u}$ in the 2D projection $Z$.
While edges can be estimated by visual information (\eg color) in Figure \ref{fig:model2} (a), these edges become susceptible to brightness changes by the 3D object's motions.
To address this, we use the depth map (Figure~\ref{fig:model2}(b)), measured by vertical distance between the viewpoint and the 3D mesh.
To decide edges based on the depth map, we employed Gaussian adaptive thresholding \cite{szeliski2022computer}, which is more responsive to local depth variations than global thresholding (\eg Canny \cite{canny1986computational}).
For each pixel $[x,y]$, an adaptive threshold $T[x,y]$ classifies edges as:
\begin{equation}
d[x, y] = 
\begin{cases} 
      1 & \text{if } D[x, y] > T[x, y]\\
      0 & \text{otherwise},
   \end{cases}
\end{equation}
producing a depth-based edge map $d$ (1 for edges, 0 otherwise). 
Thus, we define the unoccluded edge as $e^{u} = d$.
$D[x,y]$ is the depth at position $[x,y]$.
The threshold $T[x,y]$ is adaptively adjusted based on surrounding depth values within a specified window, assigning greater weight to regions closer to $[x,y]$ using Gaussian kernel weighting as:
\begin{equation}
T[x, y] = \frac{1}{R} \sum_{[i, j] \in \text{window}}G[i, j] \cdot D[x+i, y+j],
\end{equation}
where $G[i,j] = \frac{1}{2\pi\sigma^2}\exp\!\bigl(-(i^2+j^2)/(2\sigma^2)\bigr)$ and constant $R=\sum_{[i,j]\in\text{window}}G[i,j]$. 
Window in Figure \ref{fig:model2} (b) defines position $[i, j]$ in a square of width $w$ centered at $[x, y]$.
\subsection{Flow-based edge detection}
Flow-based edge detection aims to capture occluded edges $e^{o}$ in the projection $Z$.
As shown in Figure \ref{fig:model2} (a), depth-based edge detection alone can fail when body parts overlap (\eg an arm over a torso) because their depths are too similar, causing flickering.
To this end, as shown in Figure \ref{fig:model}, we propose flow-based edge detection that leverages optical flow between the current frame $Z_{i}$ and the previous frame $Z_{<i}$ (hereafter $Z_{i-1}$ for convenience) to recover missing edges in occluded regions.
To be specific, we obtain optical flow vector $v_{i-1}$ using optical flow estimators \cite{teed2020raft,lucas1981iterative} between the $Z_{i-1}$ and $Z_{i}$ in Figure \ref{fig:model3} (a).
This vector $v_{i-1}$ is a 2-dimensional vector connecting corresponding points between $Z_{i-1}$ and $Z_{i}$ to estimate optical flow.
As shown in Figure \ref{fig:model3} (b), we define these corresponding points of $Z_{i-1}$ that lie on the edge\footnote{To focus on the problem of predicting new edges based on existing ones, we assume that the chosen edges $d_{i-1}$ are well-constructed without occlusion. Occlusion-free frames can be identified by checking the overlaps between joint points of 3D coordinate space or can be automatically detected using deterministic retrievals \cite{yoon2022selective,yoon2023scanet,yoon2023counterfactual,yoon2021weakly,ma2020vlanet,luu2024predictive}.} $d_{i-1}$ as source points $p_{i-1}$ (\ie blue points).
To estimate the positions of the source points in the $i$-th frame, we shift them by $v_{i-1}$, computed as $p_{i}^{*} = p_{i-1} + v_{i-1}$.
As we focus on finding undetected edges by occlusion, as shown in Figure \ref{fig:model3} (c), we retain only $p_{i}^{*}$ that lie within the interior of $d_{i}$.
Finally, the remaining points are interpolated to construct flow-based edge $f$ in Figure \ref{fig:model3} (d).
This can be formulated on $i$-th frame as:
\begin{equation}
f_{i} = h \cdot g(p^{*}_{i},d_{i}),
\end{equation}
where the $g$ is a filter selecting valid points inside the depth-based edge $d_{i}$, and $h$ is an interpolator constructing edge from points (we use dilation for efficiency).
Therefore, we finally define the occluded edge as $e^o = f$.
\begin{figure}[h!]
  \centering
   \includegraphics[width=\linewidth]{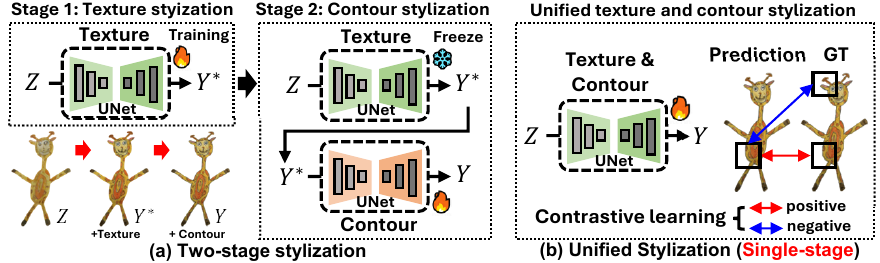}

   \caption{(a) two-stage stylization framework and (b) our single-stage unified stylization with edge-guided contrastive learning.}
   \label{fig:ssd}
   \vskip -0.2in
\end{figure}
\subsection{Unified Stylization Network}
Since multiple drawing images for various poses of an object are not always available, we assume stylization operates in the most general scenario, where each object provides only a single drawing image for training.
Figure \ref{fig:ssd} (a) shows the existing two-stage method \cite{zhou2024drawingspinup}, where texture stylization produces an intermediate output $Y^{*}$ from $Z$, followed by contour stylization to produce final $Y$.
Although this method improves contour training despite the limited contour (compared to texture) in the drawing, it uses extra labels (ground-truth $Y^{*}$) and increases both memory and time costs.
To address these, Figure \ref{fig:ssd} (b) shows our unified (single-stage) stylization network (USNet), which learns all drawing properties at once.
As contours are relatively sparse compared to textures, we introduce edge-guided contrastive learning for more discriminative contour stylization.
To be specific, we first apply patch-wise reconstruction loss for stylization by predicting original drawing-style input frame $X \in \mathbb{R}^{N \times 3}$ from the 2D projection frame $Z \in \mathbb{R}^{N \times 3}$ as $\mathcal{L}_{recon} = \sum_{i=1}^{N} (Y_{i} - X_{i})^{2}$, where $Y$ is the stylization output in Equation (\ref{eq:stylenet}), $N$ is the number of patches, and each patch has 3 channels (RGB).
We then extend this to contrastive learning, enforcing an inequality as:
\begin{equation}
\text{cos}(Y_{j}, X_{j}) > \text{cos}(Y_{j}, X_{k}), \quad \forall k \neq j,
\end{equation}
so that the cosine similarity between $Y_{j}$ and its ground-truth $X_{j}$ for contour patches exceeds that of $Y_{j}$ with other patches $X_{k}$ as hard negatives. 
The $X_{k}$ are randomly sampled from patches containing the occlusion-robust edge $e$.
This inequality is ensured by contrastive ranking loss as below:
\begin{equation}
\mathcal{L}_{c} = \sum\nolimits_j \max \left( 0, \text{cos}(Y_{j}, X_{k}) - \text{cos}(Y_{j}, X_{j}) + \delta \right),
\end{equation}
where $\delta=0.1$ maintains a margin. Our final loss for USNet combines the two terms as 
$\mathcal{L} = \mathcal{L}_{recon} + \mathcal{L}_{c}$.

%% file: sec/3_experiment.tex
\section{Experiment}
\subsection{Experimental Settings}
\paragraph{Implementation Details.}
We use RAFT \cite{teed2020raft} for optical flow.
Hyperparameters were chosen by study in Table \ref{tab:ablation}, confirming $w=9$ and dilation-based interpolation for $h$.
For image-to-3D diffusion, we use Wonder3D \cite{long2024wonder3d} and Adobe Mixamo\footnote{we use online platform (https://www.mixamo.com), but other automatic joint estimation methods \cite{hong2023joint,choi2021beyond} can also be applicable.} skeleton of 65 joints for automatic rigging.
\paragraph{Data and Baselines.}
We selected 120 test characters from the Amateur Drawings \cite{10.1145/3592788}, following the same set in \cite{zhou2024drawingspinup}.
Training (fine-tuning) uses a single pose given by the characters, while inference applies stylization to various poses.
For inference, each character is given 20 non-occluded motions (non-occlusion set) and 20 occluded motions (occlusion set)\footnote{Our supplemental material provides all Mixamo motion names} for a total of 4800 samples.
We also prepare 80 separate characters for a validation set to investigate the effectiveness of occlusion.
The 3D animation system includes non-stylization (DreamGaussian~\cite{tang2023dreamgaussian}, Wonder3D~\cite{long2024wonder3d}) and stylization methods (DrawingSpinUp~\cite{zhou2024drawingspinup}, OSF).
\subsection{Evaluation Metrics}
We evaluate each drawing animation in terms of overall texture and contour by separating the two.
For texture quality, we measure temporal consistency (via CLIP \cite{radford2021learning} similarity, SSIM \cite{wang2004image} for smoothness, and FID \cite{heusel2017gans} for naturalness) and fidelity (via LPIPS \cite{zhang2018unreasonable} to assess perceptual preservation of object identity).
We construct ground-truth distributions using input drawings, and then measure FID by comparing distributions of the generated drawings.
For contour quality, we extract contours using the method in \cite{zhou2024drawingspinup} and evaluate their temporal consistency with CLIP, SSIM, and FID.
All metrics are averaged over 10 runs with different seeds and Human evaluation of preferences is performed.
\begin{figure*}[t]
  \centering
   \includegraphics[width=\linewidth]{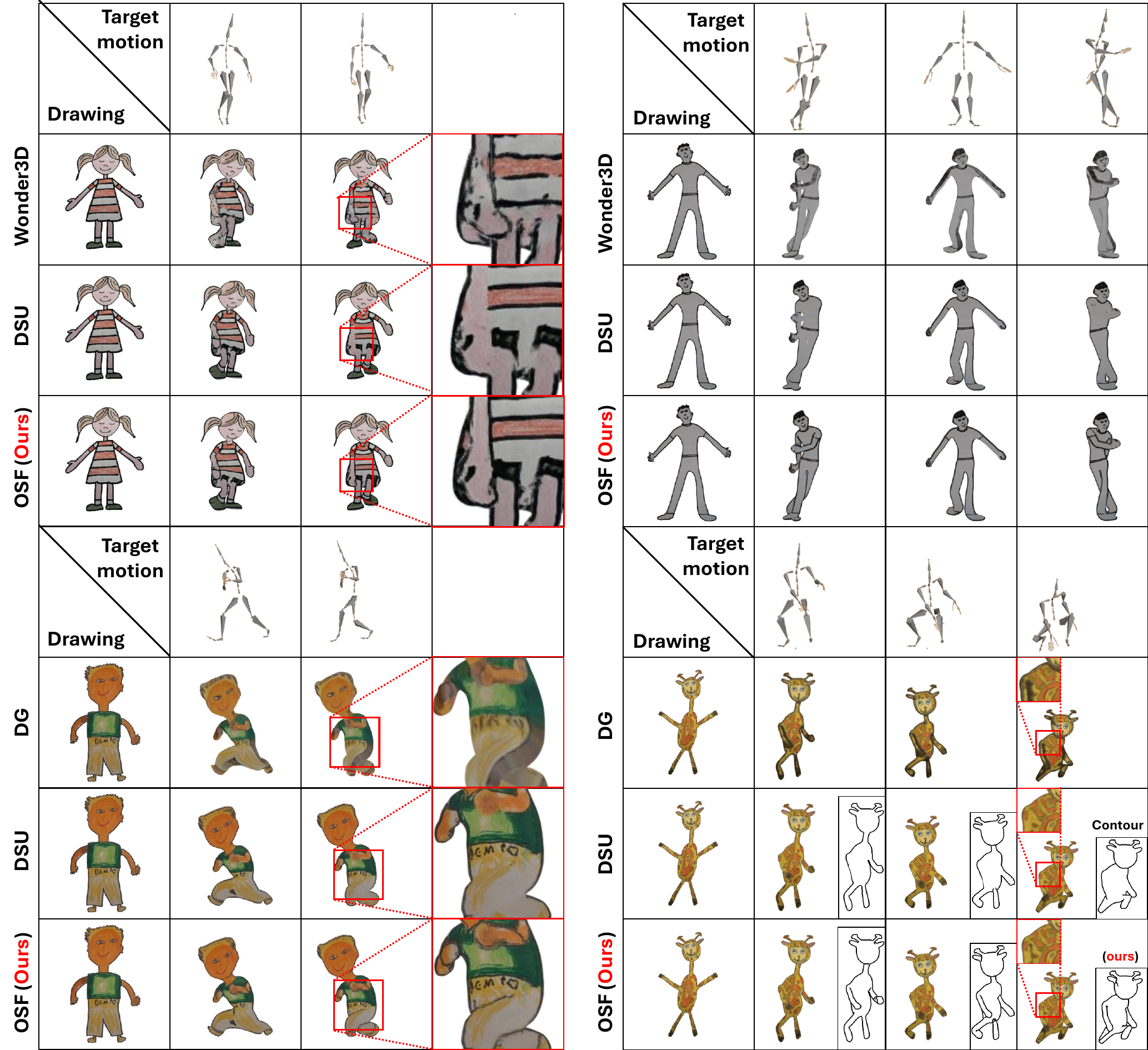}

   \caption{Qualitative comparisons across different drawing-based 3D animation models. The red box highlights a zoomed view of the stylization applied to animated drawing objects. The contour in the bottom right is extracted by the contour estimator from \cite{zhou2024drawingspinup}. Appendix also provides more qualitative results. DSU: DrawingSpinUp, DG: DreamGaussian.}
   \label{fig:qual}
   \vskip -0.2in
\end{figure*}
\begin{table*}[t]
\small
  \caption{Quantitative evaluations on recent drawing-based 3D animation models, reported in a format of (non-occlusion set / occlusion set). DSU: DrawingSpinUp, USNet: Unified Stylization Network, FDED: Flow-Depth Edge Detection. (OSF = USNet + FDED).}
  \centering
  \begin{tabular}{l ccccc c cc c}
    \toprule
            \multirow{2}{*}{Method} &\multicolumn{4}{c}{Texture} & &\multicolumn{3}{c}{Contour} &\multirow{2}{*}{Human} \\ \cline{2-5} \cline{7-9}
		  & CLIP ↑ & SSIM ↑ & FID ↓  & LPIPS ↓ && CLIP ↑ &SSIM ↑ & FID ↓ \\ 
\midrule
    DreamGaussian \cite{tang2023dreamgaussian} & 0.903/0.898 & 0.842/0.813 & 554/592 & 0.725/0.798 && 0.894/0.875 & 0.828/0.786 & 441/513 &0.01 \\ \hline
    Wonder3D \cite{long2024wonder3d}   & 0.914/0.891 & 0.844/0.816 & 531/586 & 0.714/0.782 && 0.906/0.886 & 0.832/0.801 & 432/476 &0.03 \\ \hline
    DSU \cite{zhou2024drawingspinup}          & 0.958/0.927 & 0.889/0.842 & 313/368 & 0.629/0.692 && 0.941/0.919 & 0.879/0.830 & 215/240 &0.35 \\
    DSU + FDED (Ours)     & 0.971/0.962 & 0.903/0.881 & 297/331 & 0.601/0.637 && 0.978/0.974 & 0.936/0.910 & 191/199 & - \\ \hline
    USNet (Ours)       & 0.963/0.936 & 0.896/0.851 & 308/361 & 0.612/0.685 && 0.946/0.925 & 0.882/0.832 & 212/236 & - \\
    USNet + FDED (Ours)  & \textbf{0.974}/\textbf{0.963} & \textbf{0.910}/\textbf{0.888} & \textbf{293}/\textbf{325} & \textbf{0.586}/\textbf{0.622} && \textbf{0.982}/\textbf{0.977} & \textbf{0.938}/\textbf{0.914} & \textbf{188}/\textbf{195} & 0.61 \\
    \bottomrule
  \end{tabular}
  \label{tab:1}
\end{table*}
\subsection{Experimental Results}
\begin{figure}[t]
  \centering
   \includegraphics[width=\linewidth]{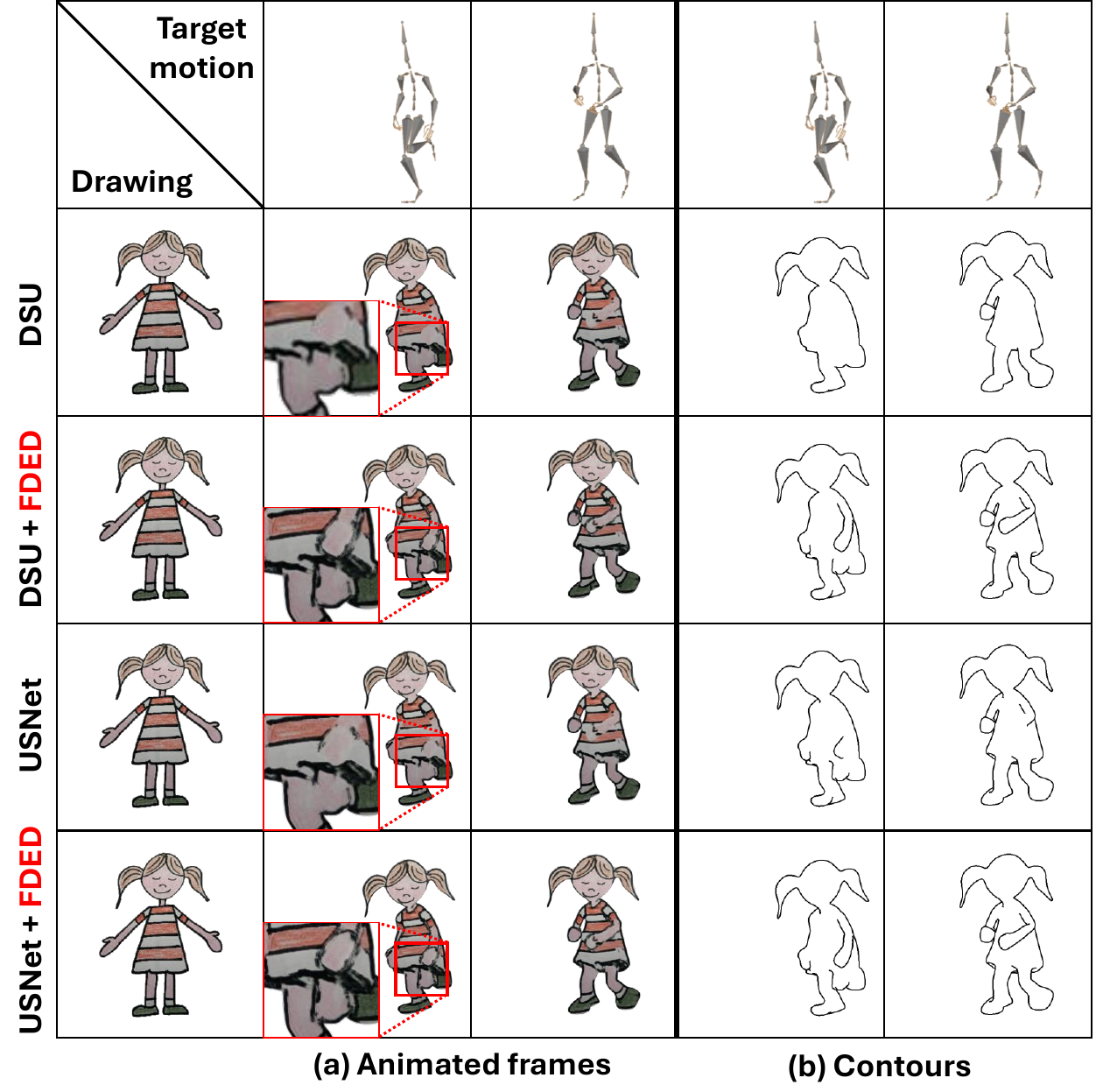}

   \caption{Effectiveness of FDED on stylization models, showing (a) resulting frames and (b) their contours with and without FDED.}
   \label{fig:effectiveness}
   \vskip -0.2in
\end{figure}
\paragraph{Qualitative Comparisons.}
Figure \ref{fig:qual} presents the 3D animation results of drawings across different 3D animation systems.
Our comparison evaluates dynamic motions, such as dancing and sitting, applied to a variety of character drawings, ranging from humans to animals.
Wonder3D \cite{long2024wonder3d} and DreamGaussian \cite{tang2023dreamgaussian} are image-to-3D models, applied to drawing-based 3D animation task without including a stylization process.
Without a stylization process, these models exhibit distortions in the drawing style, such as thickened contours and blurred stroke patterns.
DrawingSpinUp (DSU) \cite{zhou2024drawingspinup} and our proposed Occlusion-robust Stylization Framework (OSF) are stylization-based models designed for 3D animation of drawings, performing stylization to preserve the original style of the input image.
Both models retain the style of the input image well. 
However, DSU tends to lose contour details in occluded body parts (\ie red box), leading to flickering. 
This issue is particularly evident in the contour, as shown in the box visualizing extracted giraffe's contours in bottom-right of Figure \ref{fig:qual}.
OSF demonstrates robust contour preservation, even in the presence of occlusions occurring during various motions.
\paragraph{Quantitative Results.}
Table \ref{tab:1} presents quantitative results on different animation models.
We evaluated the animation quality on motions without occlusion (\ie non-occlusion set) and with occlusion (\ie occlusion set).
As contours are a key attribute defining an object's shape, we extract them for more detailed evaluation.
The proposed FDED can serve as an additional condition for any stylization model, such that we also integrate it into DSU, observing steady gains in both stylization baselines (DSU and USNet).
With the application of FDED, both consistency and fidelity have improved, with consistency enhancements largely attributed to improved contour stability.
Our proposed USNet resolves the complex two-stage inference structure of DSU, while its performance is also comparable to or better than DSU due to the edge-guided contrastive learning.
The effectiveness of the contrastive learning is also validated in Figure \ref{fig:edge_guided_con} about learning optimization and qualitative comparisons.
\begin{table}[t]
\caption{Ablation study on the main modules of FDED and their hyperparameters under validation occlusion set (D: depth-based edge, F: flow-based edge, $w$: window size of adaptive thresholding, $h$: edge reconstruction, and spline: B-spline interpolation).}
\centering
\begin{tabular}{lcccc}
    \toprule
    \multirow{2}{*}{Method} & \multicolumn{2}{c}{Texture} & \multicolumn{2}{c}{Contour} \\ \cline{2-5}
    & CLIP↑  & SSIM↑ & CLIP↑ & SSIM↑ \\
    \midrule
    w/o FDED & 0.946 & 0.873 & 0.929 & 0.849   \\
    w/ FDED (D) & 0.956 & 0.887 & 0.941 & 0.876   \\ 
    w/ FDED (D + F) & \textbf{0.971} & \textbf{0.906} & \textbf{0.980} & \textbf{0.934}  \\ \midrule
    D + F ($w$: 7) & 0.968 & 0.905 & 0.977 & 0.931  \\
    D + F ($w$: 9) & \textbf{0.971} & \textbf{0.906} & \textbf{0.980} & \textbf{0.934} \\ 
    D + F ($w$: 11) & 0.970 & 0.904 & 0.978 & 0.932  \\
    D + F ($w$: 13) & 0.968 & 0.903 & 0.977 & 0.930  \\
    \midrule
    D + F ($h$: spline) & 0.969 & 0.906 & 0.977 & 0.930 \\
    D + F ($h$: dilation) & \textbf{0.971} & \textbf{0.906} & \textbf{0.980} & \textbf{0.934}  \\
    \bottomrule
\end{tabular}
\label{tab:ablation}
\vskip -0.1in
\end{table}
\subsection{Ablation Study}
\begin{figure}[t]
  \centering
   \includegraphics[width=\linewidth]{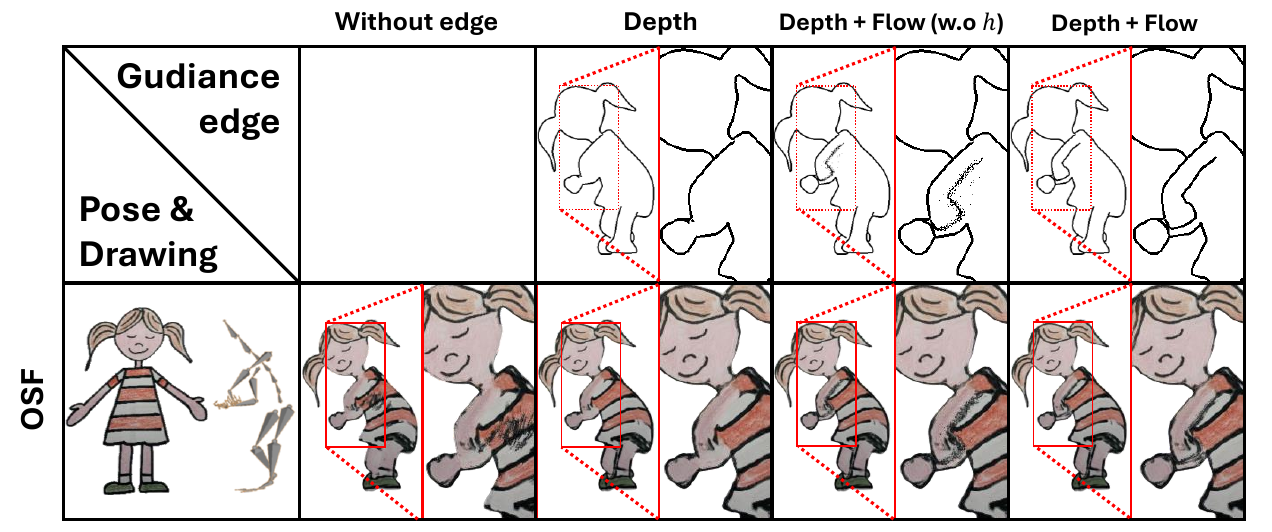}

   \caption{Ablation study about different edges in FDED. The red box shows the zoomed results in the occluded areas. The $h$ denotes edge interpolation in flow-based edge detections.}
   \label{fig:ablation_on_edge}
   \vskip -0.1 in
\end{figure}

\begin{figure}[t]
  \centering
   \includegraphics[width=\linewidth]{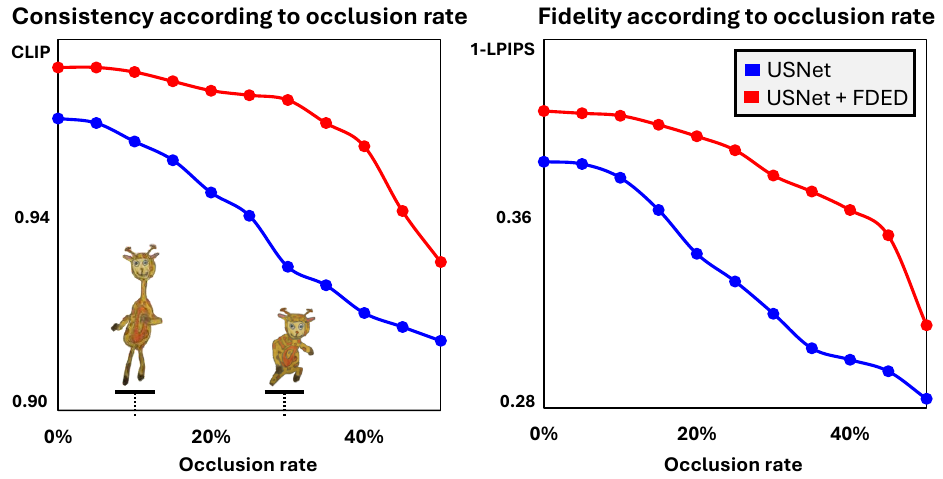}

   \caption{Robustness analysis of FDED on stylization-based 3D animation model (USNet). The characters illustrate examples of varying occlusion levels based on occlusion rate.}
   \label{fig:robustness}
   \vskip -0.1in
\end{figure}

\paragraph{Effectiveness of FDED.}
Figure \ref{fig:effectiveness} demonstrates the effectiveness by applying FDED to stylization-based animation models (DSU and USNet).
Without FDED (Figure~\ref{fig:effectiveness} (b)), both struggle to render contours in occluded regions accurately, whereas FDED-integrated versions preserve contours more clearly.
Table \ref{tab:ablation} shows an ablation study on depth-based edge (for unoccluded edge) and flow-based edge (for occluded edge), along with hyperparameter variations.
From the first section, providing guidance edges significantly boosts performance, especially with flow-based edges, revealing the vulnerability of existing methods to occlusion.
\begin{figure}[t]
  \centering
   \includegraphics[width=\linewidth]{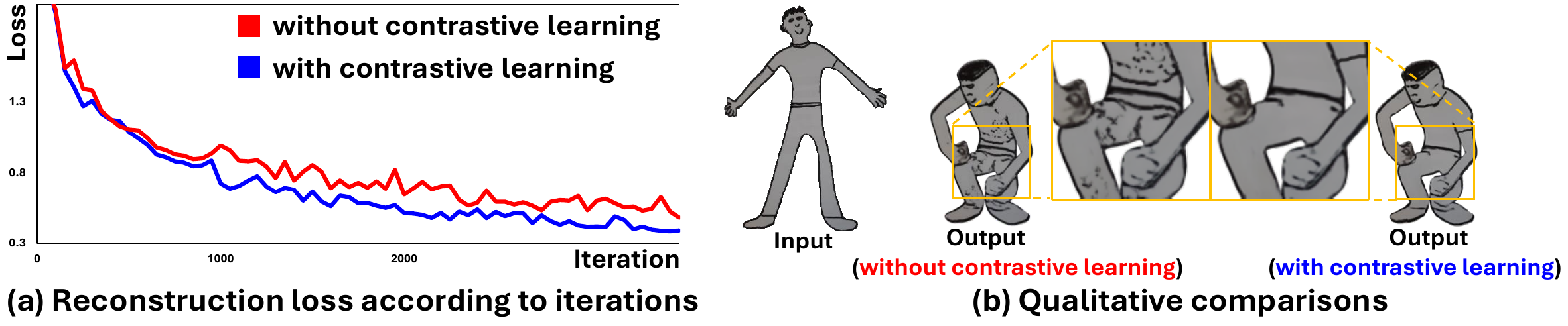}

   \caption{Ablation study on edge-guided contrastive learning. (a) demonstrates enhanced reconstruction loss, integrated with our contrastive learning and (b) shows qualitative comparisons of this.}
   \label{fig:edge_guided_con}
   \vskip -0.2 in
\end{figure}
This improvement mainly stems from enhanced contour quality.
We also present qualitative results of this according to different guidance edges in Figure \ref{fig:ablation_on_edge}.
In the second section, we also examine changes in the window size for Gaussian adaptive thresholding used in depth-based edge detection and find it to be relatively insensitive, presumably due to the continuous nature of the drawing.
The third section covers two approaches for flow-based edges: B-spline interpolation \cite{de1978practical} and dilation. 
Dilation produces better results, as it mitigates the scarcity of source points.
\paragraph{Effectiveness of contrastive learning.}
Figure \ref{fig:edge_guided_con} shows the effectiveness of edge-guided contrastive learning in terms of reconstruction loss and outcomes.
Applying contrastive learning leads to a faster drop in reconstruction loss with fewer iterations (Figure \ref{fig:edge_guided_con} (a)), ultimately converging to a lower value.
Consequently, the stylization provides sharper textures and contours, shown in Figure \ref{fig:edge_guided_con} (b).
\begin{figure}[t]
  \centering
   \includegraphics[width=\linewidth]{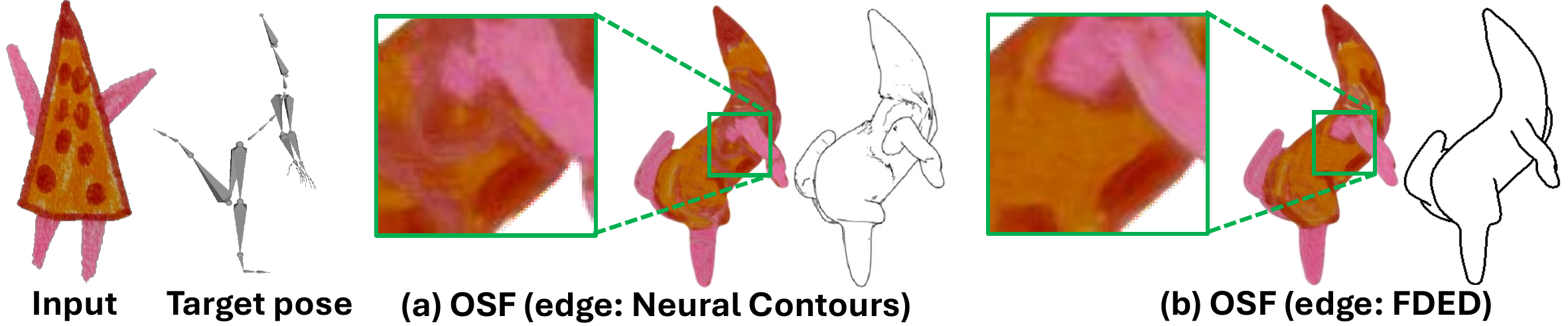}

   \caption{Comparison results about OSF with (a) 3D edge detection \cite{liu2020neural} and (b) Flow-Depth Edge Detection (FDED, Ours).}
   \label{fig:ablation_on_other_edge}
   \vskip -0.2in
\end{figure}
\paragraph{Ablation study on edge guidance.}
Figure \ref{fig:ablation_on_edge} shows qualitative animation results using various edges obtained from FDED. With USNet as the base stylization network, we test four edge configurations: (1) none, (2) depth-only, (3) depth + flow (no edge interpolation), and (4) depth + flow (with interpolation).
Without edge guidance, stylization is highly unstable, producing blurred contours and strokes. We highlight occluded regions in a zoomed view (red box).
Depth-only edges capture the overall silhouette but miss contours in occluded regions, causing incomplete animations. Incorporating flow-based edges addresses occluded contours but appears scattered when interpolation is absent. By adding the edge interpolation, flow-based edges produce sharper contours, improving the final result.
Figure \ref{fig:ablation_on_other_edge} presents the comparison results of 3D edge detection \cite{liu2020neural}.
Although the drawn 3D edges convey finer detail and three-dimensional depth, they were unsuitable for stylization guidance. 
Because stylization misinterprets them as contours, the resulting texture merges with them, causing a blurry appearance.
\paragraph{Robustness analysis on occlusion.}
Figure \ref{fig:robustness} presents the occlusion robustness evaluation of the stylization-based 3D animation system, comparing results with and without the application of FDED.
%
%
We define occlusion rate by calculating a reduction in the object's visible area\footnote{Samples only with visible occlusion are considered} compared to the original input drawing pose.
The giraffe character in Figure \ref{fig:robustness} illustrates example poses at varying occlusion levels.
As occlusion increases, the output animation quality (\ie blue) of the stylization system significantly deteriorates.
Applying FDED to the system enhances robustness, extending the range that ensures high-quality output in both consistency and fidelity.
However, with occlusion above 40\%, quality deteriorates, due to the complexity of multiple occluded areas, obscuring contours and strokes.
For this, quantization \cite{luu2024mitigating} into patch-wise encoding may enhance robustness.
%
\begin{table}[t]
\caption{Inference time and memory usage across stylization networks (excluding 3D diffusion model for isolated analysis).}
\centering
  \begin{tabular}{lcc}\toprule  
Method & seconds/frame & memory \\\midrule
DSU \cite{zhou2024drawingspinup}  & 0.276 & 11.62 GB\\
DSU \cite{zhou2024drawingspinup} + FDED (Ours) & 0.282 &  11.91 GB\\ \midrule
USNet (Ours)  & 0.108 & 4.81 GB\\
USNet + FDED (Ours) & 0.115 & 5.05 GB\\ \bottomrule
\end{tabular}
\label{tab:complexity}
\vskip -0.15in
\end{table}
\paragraph{Computational Complexity Analysis.}
\vskip -0.15in
Table \ref{tab:complexity} presents a computational analysis of stylization processing in terms of inference time and resource usage for the DSU and USNet baselines.
The proposed USNet resolves the DSU's two-stage complex inference process into a single-stage process, achieving over twice the efficiency in inference time and memory usage.
Furthermore, the integration of FDED introduces minimal overhead in inference time.
This can be improved with diffusion acceleration techniques \cite{koo2024wavelet,huang2022fastdiff}.
%
\section{Conclusion}
This paper addresses the challenge of generating 3D animations from hand-drawn images while preserving stylistic details such as rough contours and strokes. Existing models suffer from quality deterioration, especially under occlusion, due to a stylization pose gap between the unoccluded poses used in a training and the dynamic poses encountered in an inference. To this end, we introduce Occlusion-robust Stylization Framework (OSF), which improves robustness of stylization under occlusion via our designed flow-depth edge detection. Furthermore, OSF operates in a single pass, achieving faster inference and reducing memory usage.
\clearpage
\section*{Acknowledgements}
This work was partly supported by Institute for Information \& communications Technology Planning \& Evaluation (IITP) grant funded by the Korea government(MSIT) (No.RS-2021-II211381, Development of Causal AI through Video Understanding and Reinforcement Learning, and Its Applications to Real Environments) and partly supported by Institute of Information \& communications Technology Planning \& Evaluation (IITP) grant funded by the Korea government(MSIT) (No. RS-2022-II0951, Development of Uncertainty-Aware Agents Learning by Asking Questions)